\documentclass[12pt]{iopart}

\usepackage{graphicx}
\usepackage{iopams}
\usepackage{subfigure}
\usepackage{arydshln}

\newcommand{\bra}[1] {\langle #1 |}
\newcommand{\ket}[1] {| #1 \rangle}

\newcommand{\me}[3]{\bra{#1} #2 \ket{#3}}

\begin{document}

\title{Measuring the defect structure orientation of a single NV$^-$ centre in diamond}

\author{M W Doherty$^1$, J Michl$^2$, F Dolde$^2$, I Jakobi$^2$, P Neumann$^2$, N B Manson$^1$ and J Wrachtrup$^2$}

\address{$^1$ Laser Physics Centre, Research School of Physics and Engineering, Australian National University, Australian Capital Territory 0200, Australia.}
\ead{marcus.doherty@anu.edu.au}
\address{$^2$ 3. Physikalisches Institut, Research Center SCoPE and IQST, Universit\"at Stuttgart, Pfaffenwaldring 57. D-70550 Stuttgart, Germany.}

\begin{abstract}
The negatively charged nitrogen-vacancy (NV$^-$) centre in diamond has many exciting applications in quantum nano-metrology, including magnetometry, electrometry, thermometry and piezometry. Indeed, it is possible for a single NV$^-$ centre to measure the complete three-dimensional vector of the local electric field or the position of a single fundamental charge in ambient conditions. However, in order to achieve such vector measurements, near complete knowledge of the orientation of the centre's defect structure is required. Here, we demonstrate an optically detected magnetic resonance (ODMR) technique employing rotations of static electric and magnetic fields that precisely determines the orientation of the centre's major and minor trigonal symmetry axes. Thus, our technique is an enabler of the centre's existing vector sensing applications and also motivates new applications in multi-axis rotation sensing, NV growth characterization and diamond crystallography.
\end{abstract}

\submitto{\NJP}
\maketitle

\section{Introduction}

The negatively charged nitrogen-vacancy (NV$^-$) centre is a remarkable point defect in diamond \cite{review} that is at the frontier of quantum technology. In particular, the NV$^-$ centre has many exciting applications as a highly sensitive quantum sensor in nano-metrology, including  magnetometry \cite{mag1,mag2,mag3,mag4}, electrometry \cite{dolde11, dolde14}, thermometry \cite{toyli12,toyli13,neumann13,kucsko13,doherty14a}, piezometry \cite{doherty14}, and gyroscopy \cite{maclaurin12,ledbetter12}. High sensitivity is principally achieved by the long-lived coherence of the centre's electron spin, which persists in ambient and extreme conditions \cite{toyli12,balasubramanian09,pham12}. Nanoscale sensing is enabled by the atomic size of the NV$^-$ centre, its bright fluorescence that allows single centres to be located with nano-resolution and its mechanism of optical spin-polarization/ readout that allows the magnetic resonance of its electron spin to be optically detected \cite{review}.

Single NV$^-$ centres have demonstrated the ability for three-dimensional vector sensing of electric fields \cite{dolde11} with single fundamental charge sensitivity \cite{dolde14}. NV$^-$ electrometry is achieved by observing the spin resonances as an applied magnetic field is rotated.
Each vector component of the unknown electric field can be determined from its combined effect with the applied magnetic field, as long as the precise orientation of the magnetic field with respect to the defect structure of the NV$^-$ centre is known.
The latter requirement has not yet been fulfilled and, as a consequence, complete NV$^-$ vector electrometry has not been possible.
In principle, vector magnetic field sensing could also be achieved using a single NV$^-$ centre by an analogous technique, where an applied electric field is rotated in place of an applied magnetic field, but this is yet to be demonstrated.
Whilst vector magnetic field sensing has been demonstrated using an ensemble of at least four differently orientated NV$^-$ centres
\cite{steinert10,steinert13,le_sage13,maertz10,pham11,hall12}, a demonstration using a single NV$^-$ centre is likely to have several advantages due to the superior spin coherence of the single centre and its potential for greater spatial resolution.
Here, we describe a technique for the measurement of the defect structure orientation of a single NV$^-$ centre relative to applied electric/ magnetic fields, which is key to the realization of high-sensitivity NV$^-$ vector sensing.

More generally, this technique will also prove useful for material science because it is prototypical for measuring the defect structure of spin defects with trigonal symmetry, such as (hh) or (kk) divacancies in silicon carbide \cite{isoya08}.
The ability to precisely characterise which orientations of NV$^-$ centres are formed as the result of different fabrication methods will provide an invaluable tool to the study of the NV$^-$ growth mechanisms \cite{michl14}.
Such detailed knowledge of the growth mechanisms can be used to improve the fabrication of homogeneous ensembles of NV$^-$ centres, within which all centres are aligned and therefore can be more effectively employed in hybrid quantum devices (e.g. coupled superconducting resonators and NV$^-$ spin ensembles \cite{amsuess11}).
Measuring the structure of a single NV$^-$ centre also allows for local characterisation of the diamond lattice.
In this way, the structure of nanodiamonds or polycrystalline sectors can be non-invasively determined.

\begin{figure}[hbtp]
\begin{center}
\includegraphics[width=0.7\columnwidth] {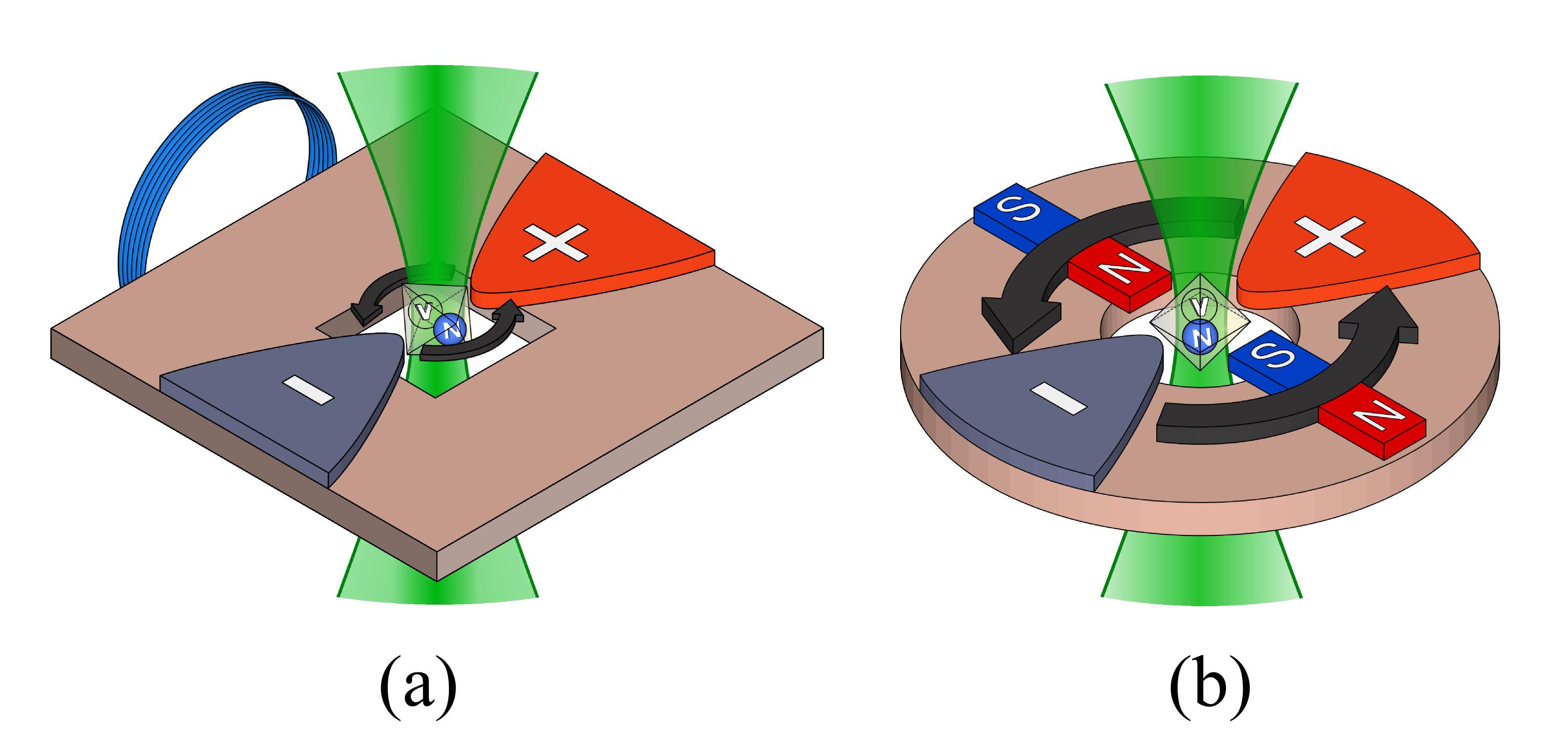}
\caption{
Two possible geometries of three-axis NV$^-$ rotation sensing:
(a) a fixed apparatus and a rotating NV$^-$ (nano-) diamond, and
(b) a fixed NV$^-$ diamond and a rotating apparatus in a micro-mechanical system.
In each geometry, the apparatus contains the necessary elements to generate electric and magnetic fields at the NV$^-$ centre.
}
\label{fig:fig2}
\end{center}
\end{figure}

The orientation measurement technique described here may be directly used to realize three-axis rotation sensing using a single NV$^-$ centre. Three-axis rotation sensing can be achieved in two different geometries (see figure \ref{fig:fig2}): (1) the NV$^-$ diamond rotates with respect to a fixed apparatus of electric and magnetic fields or (2) the apparatus rotates with respect to a fixed NV$^-$ diamond. A possible example of the first geometry is the sensing of the rotational dynamics of a nanodiamond within a biological cell \cite{mcguiness11}. An example of the second geometry is a high sensitivity rotation sensor for a micro-mechanical system, where one element of the system is free to rotate with respect to another element that contains a NV$^-$ diamond. Notably, there has been related proposals of single-axis NV$^-$ gyroscopy using single centres \cite{maclaurin12} and ensembles \cite{ledbetter12}. These proposals differ from the one made here, as they are sensitive to the rotational frequency about a single rotational axis, not the rotational coordinates of a rotation about possibly all three rotational axes.

In this paper, we demonstrate such a technique for measuring the defect structure orientation of a single NV$^-$ centre.
After further introduction to the properties of the NV$^-$ centre, we first derive the relationship between the spin resonances of a NV$^-$ centre in the presence of combined electric and magnetic fields and its defect structure.
The orientation measurement technique arises naturally form the derivation and we subsequently demonstrate the technique by measuring the defect structure orientation of a single NV$^-$ centre. The latter measurement, in essence, mimics our second proposed geometry for rotation sensing.
Based upon our demonstration, we finally examine the application of the technique to three-axis rotation sensing using single NV$^-$ centres.

\section{Theory of the orientation measurement technique}

\begin{figure}[hbtp]
\begin{center}
\includegraphics[width = 0.7\columnwidth]{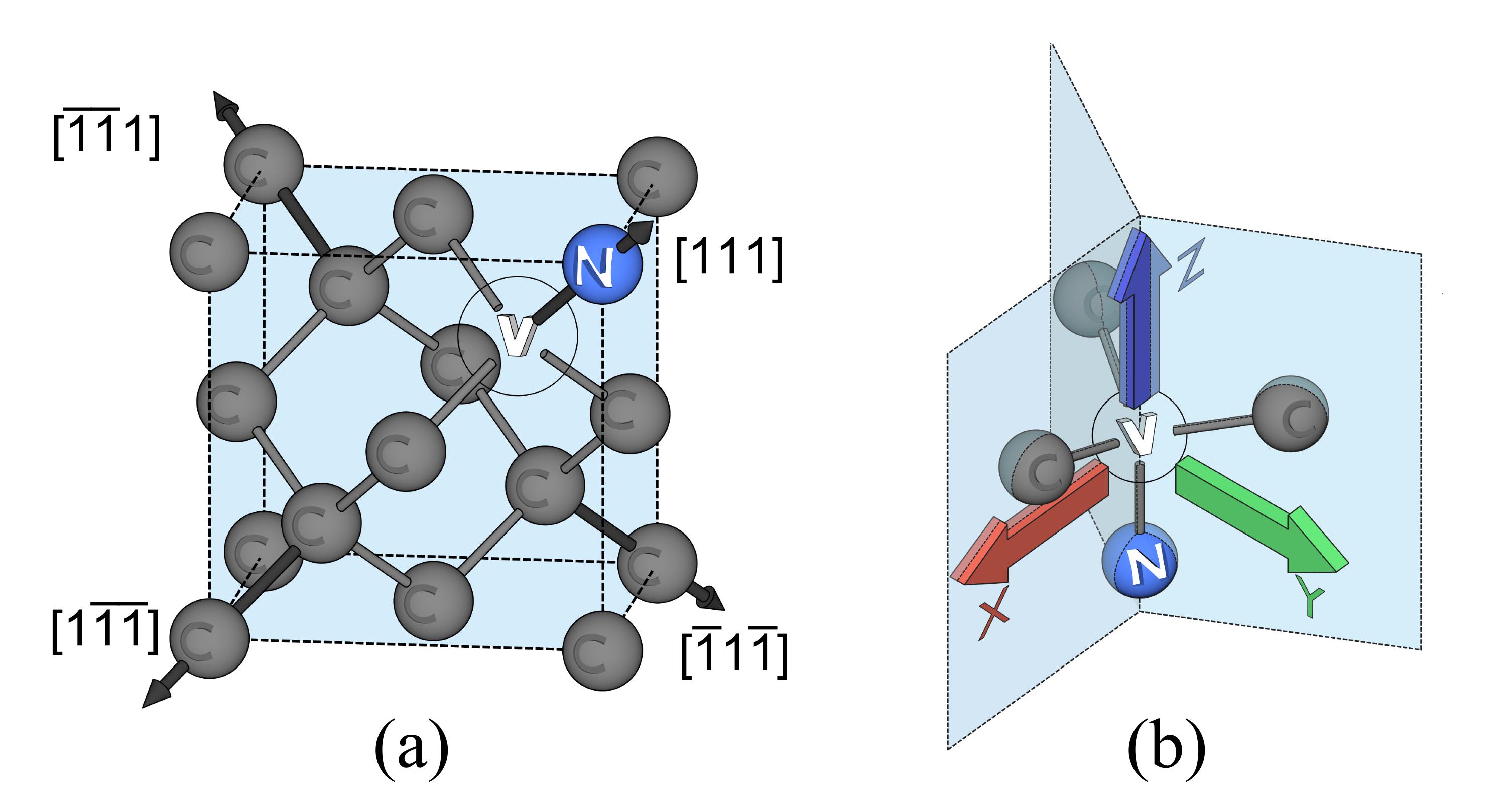}
\caption{
(a) A diamond unit cell depicting the four possible alignments of the NV defect structure's major symmetry axis.
(b) The defect structure of an NV centre, including its major trigonal, symmetry axis $z$, its reflection planes and one (of the three possible) definitions of its minor symmetry axis $x$.
}
\label{fig:fig1}
\end{center}
\end{figure}

The NV$^-$ centre is a $C_{3v}$ point defect in diamond consisting of a substitutional nitrogen atom adjacent to a carbon vacancy trapping an additional electron (refer to figure \ref{fig:fig1}a). The orientation of the centre's defect structure is characterised by the directions of its major and minor symmetry axes. The trigonal structure of the centre has a $\langle 111 \rangle$ major symmetry axis that is defined by the direction joining the nitrogen and vacancy. For a given crystal orientation, the centre's [111] major symmetry axis has four possible alignments (see figure \ref{fig:fig1}b). The centre's minor symmetry axis is defined as being orthogonal to its major symmetry axis and also contained within one of the centre's three reflection planes (eg. $[11\overline{2}]$), which corresponds to the direction joining a point on the centre's major symmetry axis and one of the vacancy's nearest neighbour carbon atoms. Considering an isolated single NV$^-$ centre, if the alignment of its major symmetry axis is known (i.e. via magnetic field alignment), then even with knowledge of the crystal orientation, the orientation of the centre's defect structure is not fully determined.

\begin{figure}[hbtp]
\begin{center}
\mbox{
\subfigure[]{\includegraphics[width=0.24\columnwidth]{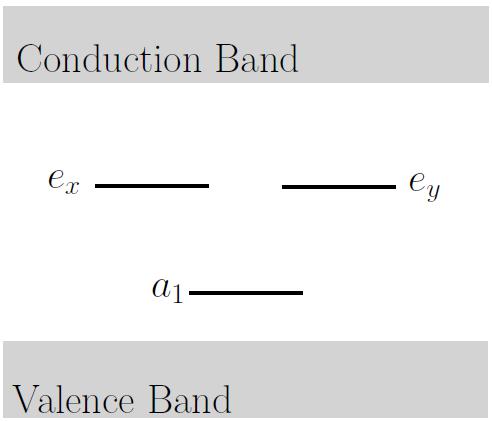}}
\subfigure[]{\includegraphics[width=0.4\columnwidth]{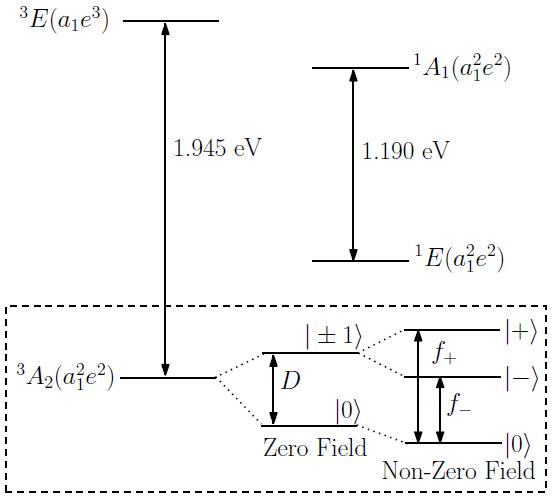}}
}
\caption{
(a) The NV$^-$ one-electron orbital level structure depicting the diamond valence and conduction bands and the three defect orbitals ($a_1$, $e_x$, and $e_y$) within the bandgap.
(b) Schematic of the center's many-electron electronic structure, including the optical 1.945 eV and infrared 1.190 eV ZPL energies.
The electronic configurations of the many-electron levels are indicated in parentheses.
Inset: The fine structure of the ground $^3A_2$ level: at zero field with a single splitting of $D\sim2.87$ GHz; and in the presence of magnetic and/ or electric fields, with a further field dependent splitting.
}
\label{fig:fig3}
\end{center}
\end{figure}

As depicted in figure \ref{fig:fig3}a, the one-electron orbital level structure of the NV$^-$ centre contains three defect orbital levels ($a_1$, $e_x$ and $e_y$). EPR observations and \textit{ab initio} calculations indicate that these defect orbitals are highly localized to the centre \cite{he93,felton09,larrson08,gali08}. Figure \ref{fig:fig3}b shows the centre's many-electron electronic structure generated by the occupation of the three defect orbitals by four electrons \cite{doherty11,maze11}, including the zero phonon line (ZPL) energies of the optical (1.945 eV/637 nm)  \cite{davies76} and infrared (1.190 eV/1042 nm) \cite{rogers08,acosta10b,manson10} transitions. The energy separations of the spin triplet and singlet levels (${^3}A_2\leftrightarrow{^1}E$ and ${^1}A_1\leftrightarrow{^3}E$) are unknown.

As depicted in the inset of figure \ref{fig:fig3}b, the ground $^3A_2$ level exhibits a zero field fine structure splitting between the $m_s=0$ and $\pm1$ spin sub-levels of $D\sim2.87$ GHz, which is principally due to first-order electron spin-spin interaction \cite{loubser78}. The spin quantization axis is thus defined by the trigonal unpaired electron spin density distribution of the ground $^3A_2$ level to be parallel to the centre's major symmetry axis. Spin-orbit and spin-spin mixing of the $^3A_2$ and $^3E$ levels makes the $^3A_2$ fine structure susceptible to electric fields, yet does not significantly perturb the g-factor of the spin magnetic interaction from its free electron value \cite{doherty12}.

A detailed derivation of the spin-Hamiltonian that describes the $^3A_2$ fine structure in the presence of electric and magnetic fields has been previously reported \cite{doherty12}. The spin-Hamiltonian derived in \cite{doherty12} is
\begin{eqnarray}
H&=&(D+ k_z E_z ) (S_z^2-2/3)+\gamma_e \vec{S}\cdot\vec{B}\nonumber \\
&&-k_x E_x (S_x^2-S_y^2)+k_y E_y (S_x S_y+S_y S_x )
\label{eq:spinhamiltonian}
\end{eqnarray}
where $\vec{S}$ are the $S=1$ dimensionless electron spin operators, $\gamma_e=g_e\mu_B/h$,  $\mu_B$ is the Bohr magneton, $g_e\sim2.003$ is the electron g-factor \cite{felton09}, $h$ is the Planck constant, $\vec{B}$ and $\vec{E}$ are the magnetic and electric fields, respectively, and $k_z=3.5(2)\,$kHz $\mu$m/V and $k_x=k_y=k_\perp=170(30)\,$kHz $\mu$m/V are the electric susceptibility parameters \cite{vanoort90}.

Although the spin-Hamiltonian derivation in \cite{doherty12} is detailed, there is limited discussion of the relationship between the spin-Hamiltonian and the defect structure of the NV$^-$ centre, in particular, the physical basis for the adopted coordinate system and the behaviour of the spin-Hamiltonian under coordinate transformations. Consequently, we present a brief review of the spin-Hamiltonian derivation in appendix A. The derivation proceeds by the application of perturbation theory to the fine structure interactions of the NV$^-$ centre and ends in the definition of each of the spin-Hamiltonian parameters in terms of integrals of the centre's orbital wavefunctions.
The key outcome of the review is the identification of the implicit coordinate system definition in (\ref{eq:spinhamiltonian}) as the choice of the $z$ and $x$ coordinate axes as being directed along the centre's major and minor symmetry axes, respectively, such that the electric susceptibility parameters are positive.
As demonstrated in appendix B, a simple molecular orbital argument can be applied to the evaluation of the orbital integrals of the electric susceptibility parameters to show that they are positive when the $z$ coordinate axis is directed from the nitrogen towards the vacancy and when the $x$ coordinate axis is directed from the $z$ coordinate axis towards one of the vacancy's three nearest-neighbour carbons (see figure \ref{fig:fig1}a). Since there are three equivalent choices of the $x$ coordinate direction due to the centre's trigonal symmetry, the spin-Hamiltonian (\ref{eq:spinhamiltonian}) is invariant under $C_3$ rotations about the major symmetry axis.

Having established the relationship between the spin-Hamiltonian and the centre's defect structure, we now discuss the orientation dependence of the centre's observable spin resonances in the presence of electric and magnetic fields. The zero-field spin states $\{\ket{0},\ket{\pm1}\}$ (defined by $S_z$ spin-projection) are mixed in the presence of electric and magnetic fields to form a new state basis $\{\ket{0},\ket{\pm}\}$ (refer to figure \ref{fig:fig3}b) \cite{doherty12}. The frequencies $f_\pm$ of the $\ket{0}\leftrightarrow\ket{\pm}$ spin transitions are \cite{dolde14}
\begin{eqnarray}
f_\pm(\vec{E},\vec{B}) & = & D+k_z E_z+3\Lambda\pm\sqrt{{\cal R}^2-\Lambda{\cal R}\sin\alpha\cos\beta+\Lambda^2}
\label{eq:spinfrequencies}
\end{eqnarray}
where $\Lambda = \gamma_e^2 B_\perp^2/2D$, ${\cal R}=\sqrt{\gamma_e^2B_z^2+k_\perp^2 E_\perp^2}$, $B_\perp=\sqrt{B_x^2+B_y^2}$, $E_\perp=\sqrt{E_x^2+E_y^2}$, $\tan\alpha=k_\perp E_\perp/\gamma_eB_z$, $\beta=2\phi_B+\phi_E$, $\tan\phi_B = B_y/B_x$ and $\tan\phi_E=E_y/E_x$. Note that the validity of the above expression is constrained by the conditions $\Lambda,{\cal R}\ll D$.

In the presence of either an electric or a magnetic field, the spin frequencies reduce to
\begin{eqnarray}
f_\pm(\vec{0},\vec{B}) & = & D+\frac{3\gamma_e^2 B^2}{2D}\sin^2\theta_B\pm\gamma_eB\cos\theta_B\sqrt{1+\frac{\gamma_e^2B^2}{4D^2}\tan^2\theta_B\sin^2\theta_B} \nonumber \\
f_\pm(\vec{E},\vec{0}) & = & D+k_z E\cos\theta_E\pm k_\perp E \sin\theta_E
\end{eqnarray}
where $E=\sqrt{E_z^2+E_\perp^2}$, $B=\sqrt{B_z^2+B_\perp^2}$, $\tan\theta_E=E_\perp/E_z$ and $\tan\theta_B=B_\perp/B_z$. In each case, the spin frequencies depend on the alignment of the electric/magnetic field with the centre's major symmetry axis (i.e. $\theta_E$ and $\theta_B$), but do not depend on the transverse orientation of the electric/ magnetic field (i.e. $\phi_E$ and $\phi_B$). This invariance to the transverse orientation of an individual field is due to the equal spin susceptibility to $x$ and $y$ field components, which is itself a consequence of the centre's trigonal symmetry.

If the electric/ magnetic field is near alignment to the centre's major symmetry axis, the spin frequencies obtain their maximum/ minimum values, such that
\begin{eqnarray}
\Delta f_\pm (\theta_B) & = & f_\pm(\vec{0},\vec{B})-D \approx \pm\gamma_eB\cos\theta_B \nonumber \\
\Delta f_\pm (\theta_E) & = & f_\pm(\vec{E},\vec{0})-D \approx k_z E\cos\theta_E
\end{eqnarray}
which may be used as a condition to identify the orientation of the symmetry axis. However, noting that $\Delta f_\pm(\theta_B+\pi)=\Delta f_\mp(\theta_B)$, it can not be determined whether the magnetic field is parallel ($\theta_B=0$) or anti-parallel ($\theta_B=\pi$) with the $z$ coordinate axis. This is not the case for the electric field, where the observable sign of the common shift $k_z E\cos\theta_E$ is directly dependent on whether the electric field is parallel or anti-parallel with the $z$ coordinate axis.
Hence, whilst either electric or magnetic fields can be used to determine the orientation of the centre's major symmetry axis, only an electric field can differentiate the precise direction of the centre's $z$ coordinate axis.

\begin{figure}[hbtp]
\begin{center}
\includegraphics[width=0.5\columnwidth]{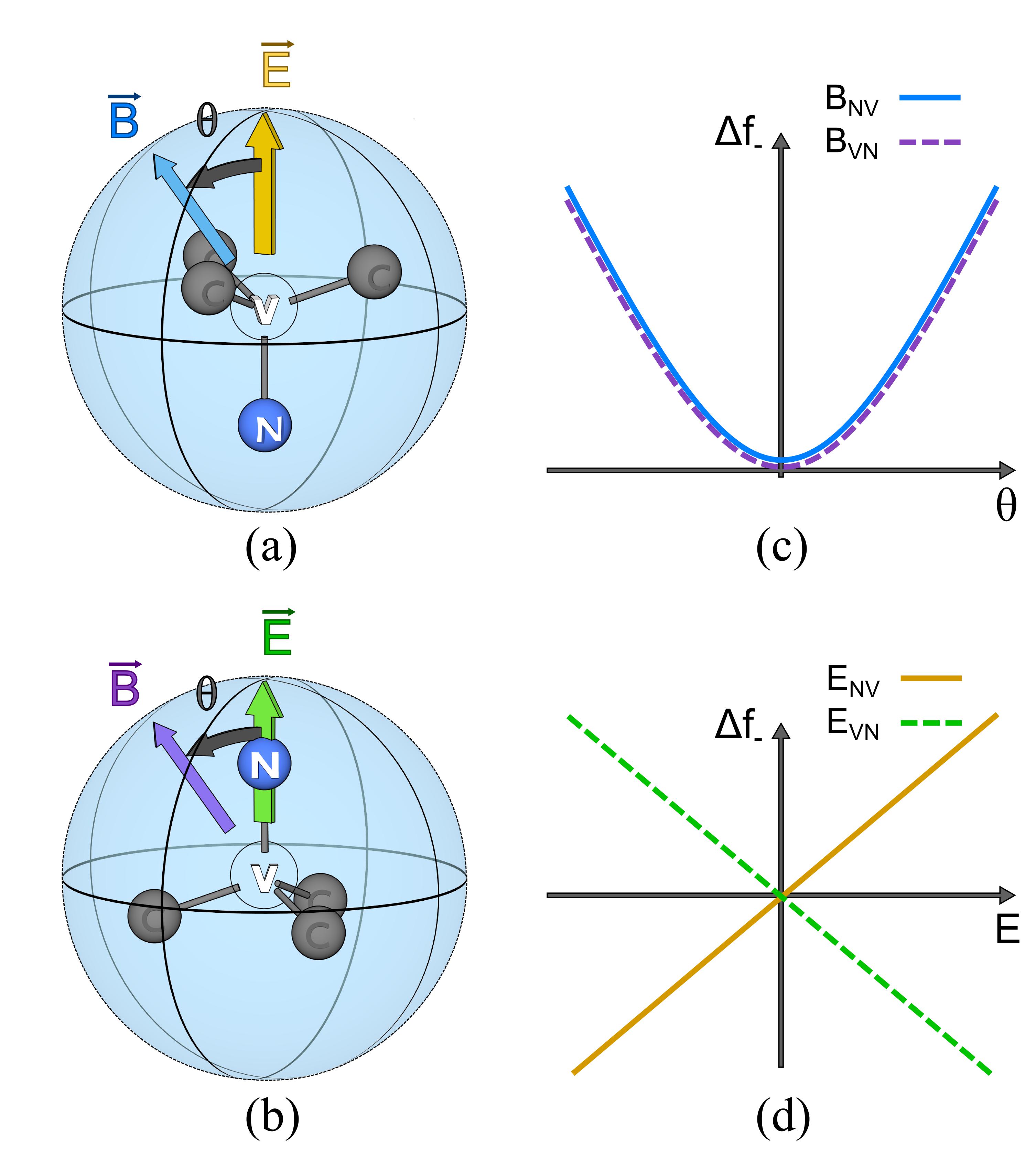}
 \caption{
Axial orientation technique.
(a,b) Magnetic and electric fields ($\vec{B}$ and $\vec{E}$) are oriented (except for misalignment angle $\theta$) parallel (a) or anti-parallel (b) to the NV major symmetry axis.
(c,d) The response of the spin transitions frequency $\Delta f_-$ does not depend on the parallel or anti-parallel orientation of the magnetic field (c) but it does depend on the electric field orientation (d).
}
\label{fig:fig4}
\end{center}
\end{figure}

This conclusion forms the basis of a technique for measuring the orientation of the $z$ coordinate axis (refer to figure \ref{fig:fig4}). Since the magnetic susceptibility of the ground state spin is much greater than its electric susceptibility, the most sensitive method is to first rotate a known magnetic field until it is aligned with the centre's major symmetry axis and then apply a parallel electric field to determine the direction of the $z$ coordinate axis. Although the electric field shift $k_z E\cos\theta_E$ is typically small, as long as it can be detected, it will be sufficient because only its sign is relevant.

It is clear from Eq.~(\ref{eq:spinfrequencies}), that the spin frequencies become dependent on the transverse orientations of electric and magnetic fields if they are simultaneously applied. This dependence obtains a maximum when the fields are transverse to the major symmetry axis
\begin{eqnarray}
f_\pm(\vec{E}_\perp,\vec{B}_\perp) & = & D+3\Lambda\pm\sqrt{k_\perp^2E_\perp^2-k_\perp E_\perp\Lambda\cos\phi+\Lambda^2} \nonumber \\
&\approx& D+(3\pm1)\Lambda\mp k_\perp E_\perp\cos(2\phi_B+\phi_E)
\label{eq:transversespinfreq}
\end{eqnarray}
where the approximation in the second line has been taken in the limit $\Lambda\gg k_\perp E_\perp$, which is relevant to typical experimental conditions \cite{dolde11,dolde14}. For an explanation of the transverse susceptibility see appendix A.

\begin{figure}[hbtp]
\begin{center}
\includegraphics[width=1.0\columnwidth]{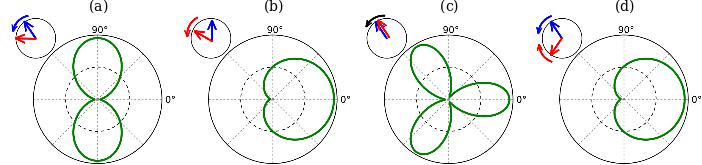}
\caption{Transverse ($B$- and $E$-) field orientation dependence of $\Delta f_-$:
(a) $\phi_B$ increased and $\phi_E=\pi$ fixed,
(b) $\phi_E$ increased and $\phi_B=\pi/2$ fixed,
(c) $\gamma$ increased and $\delta=0$ fixed, and
(d) $\delta$ increased and $\gamma=0$ fixed.
Using different values for the fixed angles would simply rotate the patterns according to equations (\ref{eq:Delta_f_pm_E_B}) and(\ref{eq:Delta_f_pm_gamma_delta}).
}
\label{fig:fig5}
\end{center}
\end{figure}

Since the second term of (\ref{eq:transversespinfreq}) leads to a splitting of the $f_\pm$ spin frequencies that is independent of the transverse field orientations, the variations of the spin frequencies as the fields are rotated may be individually observed and are governed by the third term, such that
\begin{eqnarray}
\Delta f_\pm(\phi_B,\phi_E) & \approx & f_\pm(\vec{E}_\perp,\vec{B}_\perp)-D-(3\pm1)\Lambda = \mp k_\perp E_\perp \cos(2\phi_B+\phi_E)
\label{eq:Delta_f_pm_E_B}
\end{eqnarray}
Figure \ref{fig:fig5} depicts $\Delta f_-(\phi_B,\phi_E)$ for the cases where the orientation of one field is fixed and the other is rotated. As discussed in the demonstrations of NV$^-$ vector electrometry \cite{dolde11} and single charge detection \cite{dolde14}, the mixed argument of the cosine function implies that to determine the transverse orientation of an unknown electric field by observing $\Delta f_-$ as a transverse magnetic field is rotated, one  requires knowledge of the orientation of the transverse magnetic field with respect to the centre's coordinate system (i.e. $\phi_B$). The opposite is also true for the measurement of the transverse orientation of an unknown magnetic field via a rotation of a transverse electric field. This interdependence motivates the simultaneous rotation of transverse electric and magnetic fields as a means to determine the transverse orientation of the centre's coordinate system/ defect structure.

Defining the combined electric-magnetic field angles $\gamma=(\phi_B+\phi_E)/2$ and $\delta=(\phi_B-\phi_E)/2$, the orientation dependence of the spin frequencies become
\begin{eqnarray}
\Delta f_\pm(\gamma,\delta) & \approx & \mp k_\perp E_\perp \cos(3\gamma+\delta)
\label{eq:Delta_f_pm_gamma_delta}
\end{eqnarray}
If $\delta$ is fixed whilst $\gamma$ is varied by the fields being simultaneously rotated in the same direction at the same rate, the variation of $\Delta f_-$ (see figure \ref{fig:fig5}) immediately reveals the trigonal structure of the NV$^-$ centre. Indeed, if the fields are parallel, such that $\delta=0$, the three maxima of $\Delta f_-$ directly correspond to when the fields are directed along the three equivalent $x$ coordinate axes of the centre (i.e. the directions from the major symmetry axis to the vacancy's three nearest-neighbour carbons).

\begin{figure}[hbtp]
\begin{center}
\includegraphics[width=0.5\columnwidth] {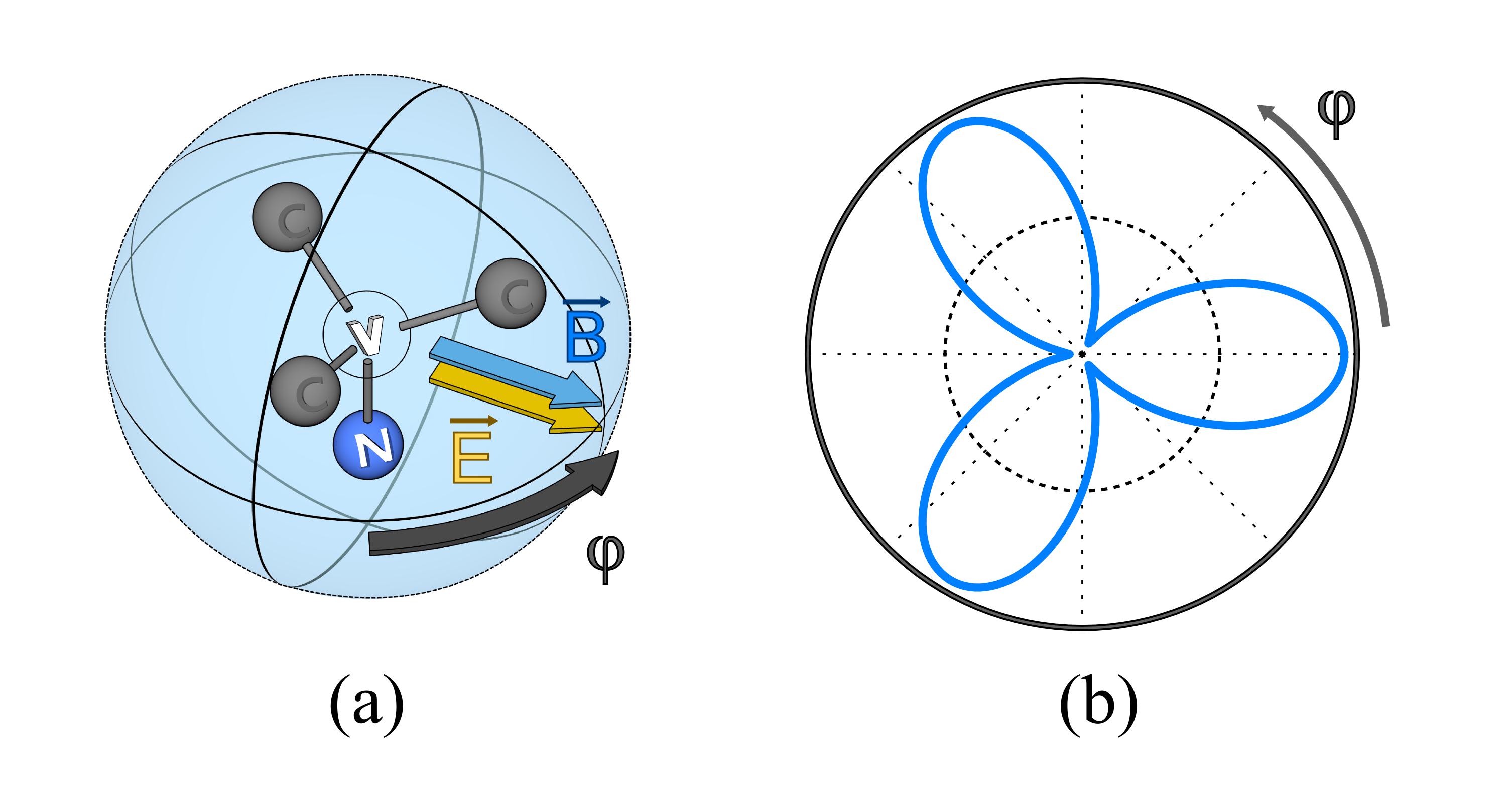}
\caption{
The transverse orientation measurement technique.
(a) Parallel electric and magnetic fields are rotated in the transverse plane whilst $\Delta f_-$ is observed.
(b) The observed trigonal pattern of $\Delta f_-$ directly corresponds to the centre's trigonal defect structure.
}
\label{fig:fig6}
\end{center}
\end{figure}

This conclusion forms the basis of a technique for measuring the orientation of the $x$ coordinate axis (refer to figure \ref{fig:fig6}).
Having first determined the orientation of the centre's major symmetry axis $z$ using the technique discussed previously, parallel electric and magnetic fields can be aligned such that they are transverse to the $z$-axis.
If the fields are rotated together as above, the direction of the $x$ coordinate axis can be identified as the field direction corresponding to one of the three maxima of $\Delta f_-$.

\section{Demonstration of the orientation measurement technique}

\begin{figure}[hbtp]
\begin{center}
\includegraphics[width=1.0\columnwidth]{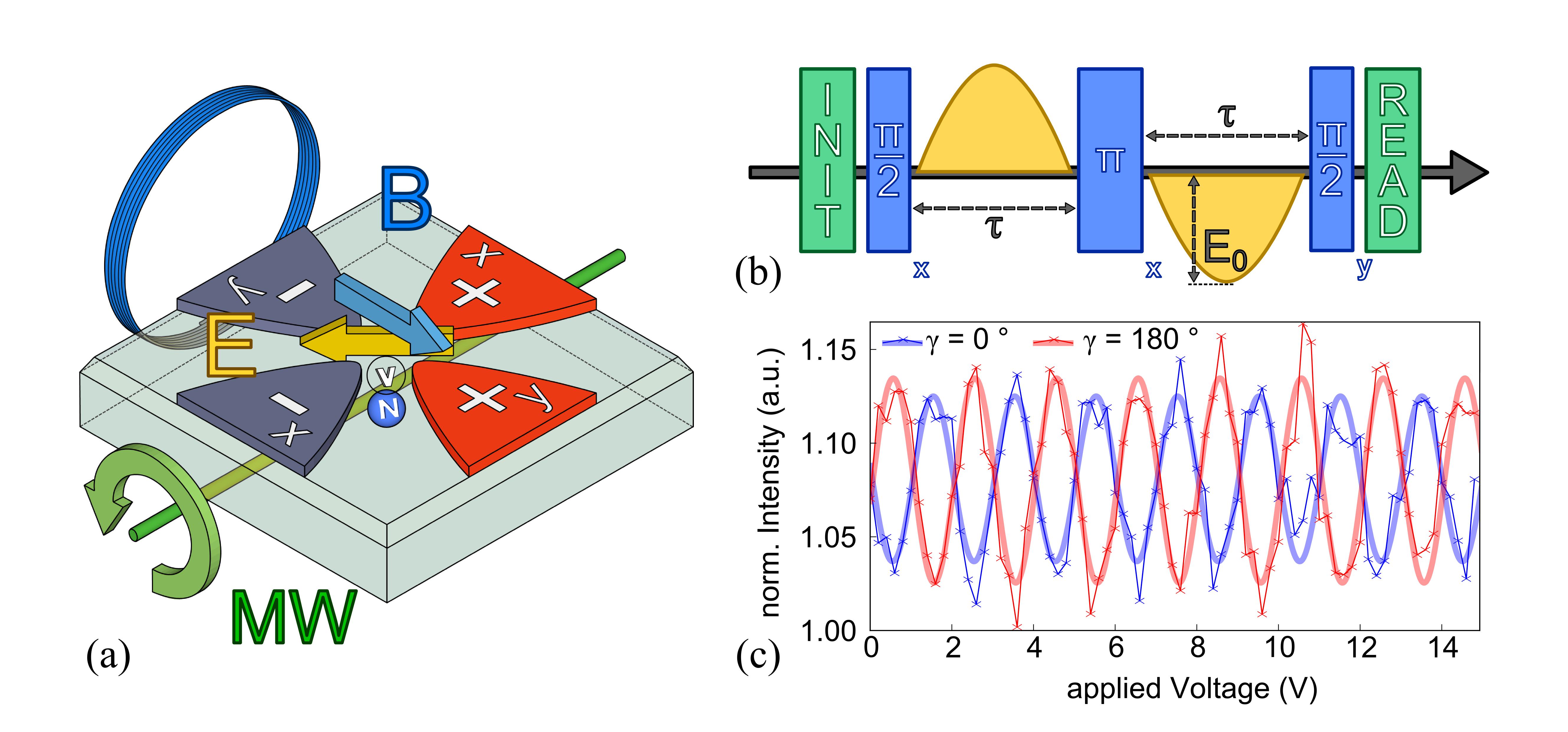}
\caption{
(a) Sketch of the experimental setup, including:
the diamond sample with an NV centre aligned perpendicular to the sample surface,
a gold quadrupole electrode microstructure for transverse electric field (yellow) generation,
a microwave wire (green) for spin manipulation and an exemplary magnetic field coil (blue).
(b) ac-electrometry ODMR sequence comprising a modified spin-echo pulse sequence interleaved by a modulated electric field $E(t)=E_0\sin\pi t/\tau$.
The spin-echo consists of two microwave $\pi/2$-pulses about orthogonal axes $x_{\mathrm{rot}}$ and $y_{\mathrm{rot}}$ of the rotating frame and an intermediate $\pi$-pulse about $x_{\mathrm{rot}}$.
The microwave pulses are separated by a delay time $\tau$.
The optical readout and preparation pulse enables the detection and reinitialization of the population difference between the spin states.
(c) Spin-echo ODMR signal over increasing electric field strength for two opposing field angles $\gamma=0,180\,\deg$ (blue, red) and equal field angle $\delta=0$ (see eq.~(\ref{eq:Delta_f_pm_gamma_delta})).
Please note that our modified spin-echo sequence is sensitive to the sign of the accumulated phase and therefore the direction of energy level shifts.
}
\label{fig:fig7}
\end{center}
\end{figure}

In this section we present a proof-of-principle demonstration of the orientation measurement technique. We performed ODMR measurements on an as-grown single NV$^-$ centre in a bulk CVD diamond sample with a $\langle111\rangle$ orientated surface. We independently verified the crystallographic directions by x-ray diffraction. However, note that this is no requirement for our method.
Our experimental setup is sketched in figure \ref{fig:fig7}a.
Three orthogonal magnetic field coils allowed a magnetic field of $\approx55\,$G to be aligned and rotated in the transverse plane.
A two-dimensional quadrupole electrode microstructure was engineered onto the diamond surface using gold lithography and similarly allowed an electric field to be aligned and rotated in the transverse plane.
Microwaves were applied to the sample by two orthogonal wires underneath the diamond sample.
Optical excitation with 532 nm light and red-shift fluorescence collection were achieved using a confocal microscope.

The dependence of $\Delta f_-$ on the orientation of the transverse electric and magnetic fields was observed using a modified spin-echo ODMR pulse sequence (see figure \ref{fig:fig7}b).
In the modified sequence, the last $\pi/2$-pulse is phase-shifted by $90^{\circ}$ to the other two pulses to enable the measurement of the sign of the accumulated phase (i.e. the direction of the energy level shift).
The microwave pulses were separated by a fixed delay time $\tau\approx 70$ $\mu$s, during which the transverse electric field was applied with a sinusoidal magnitude $E(t)=E_0\sin\pi t/\tau$.
The application of the electric field during the pulse sequence results in a net accumulated phase difference $\Phi=8\tau k_\perp E_0\cos(3\gamma+\delta)$ between $\ket{0}$ and $\ket{-}$, which is converted into an optically detectable population difference between the spin states by the final $\pi/2$-pulse.

\begin{figure}[hbtp]
\begin{center}
\includegraphics[width=0.7\columnwidth]{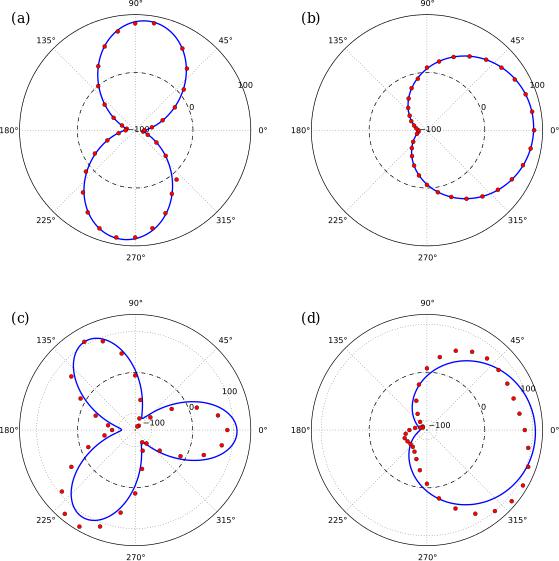}
\caption{
The observed behaviour of $\Delta f_-$ under four distinct rotation conditions of the transverse electric and magnetic fields (experimental data - points, theory - solid lines):
(a) $\phi_B$ rotated and $\phi_E=\pi$ fixed,
(b) $\phi_E$ rotated and $\phi_B=\pi/2$ fixed,
(c) $\gamma$ rotated and $\delta=0$ fixed, and
(d) $\delta$ rotated and $\gamma=0$ fixed.
Note that the ``two-leaf'' pattern of (a) differs from the ``four-leaf'' pattern observed in the previous electrometry demonstrations \cite{dolde11} because our measurements were sensitive to both the sign and magnitude of $\Delta f_-$.
}
\label{fig:fig8}
\end{center}
\end{figure}

Figure \ref{fig:fig8} depicts the observed behaviour of $\Delta f_-$ for the three distinct rotations of the transverse electric and magnetic fields.
In each case, the measurements agree well with the theoretical prediction and any small differences can be explained by small deviations of the fields from transverse alignment as they are rotated.
The strength of the measured electric field could be calculated to $E_{\mathrm{max}}=0.64\,$V$\mu$m$^{-1}$ for the maximum applied voltage of $\pm 15$ V, which compares to a simulated electric field strength of $E_{\mathrm{sim}}=0.5-0.75\,$V$\mu$m$^{-1}$ for such a quadrupole structure, depending on the exact position of the NV centre within the structure.
As discussed in the previous section, because the electric and magnetic fields are parallel ($\delta=0$), the trigonal pattern of figure \ref{fig:fig8}c directly corresponds to the orientation of the centre's trigonal defect structure and would be shifted by $180^{\circ}$ for a differently oriented NV. EPR and crystallographic results concerning the direction of the minor symmetry axis coincide within the error margins.

\section{Discussion of applications in multi-axis rotation sensing}

\begin{figure}[hbtp]
\begin{center}
\includegraphics[width=0.6\columnwidth] {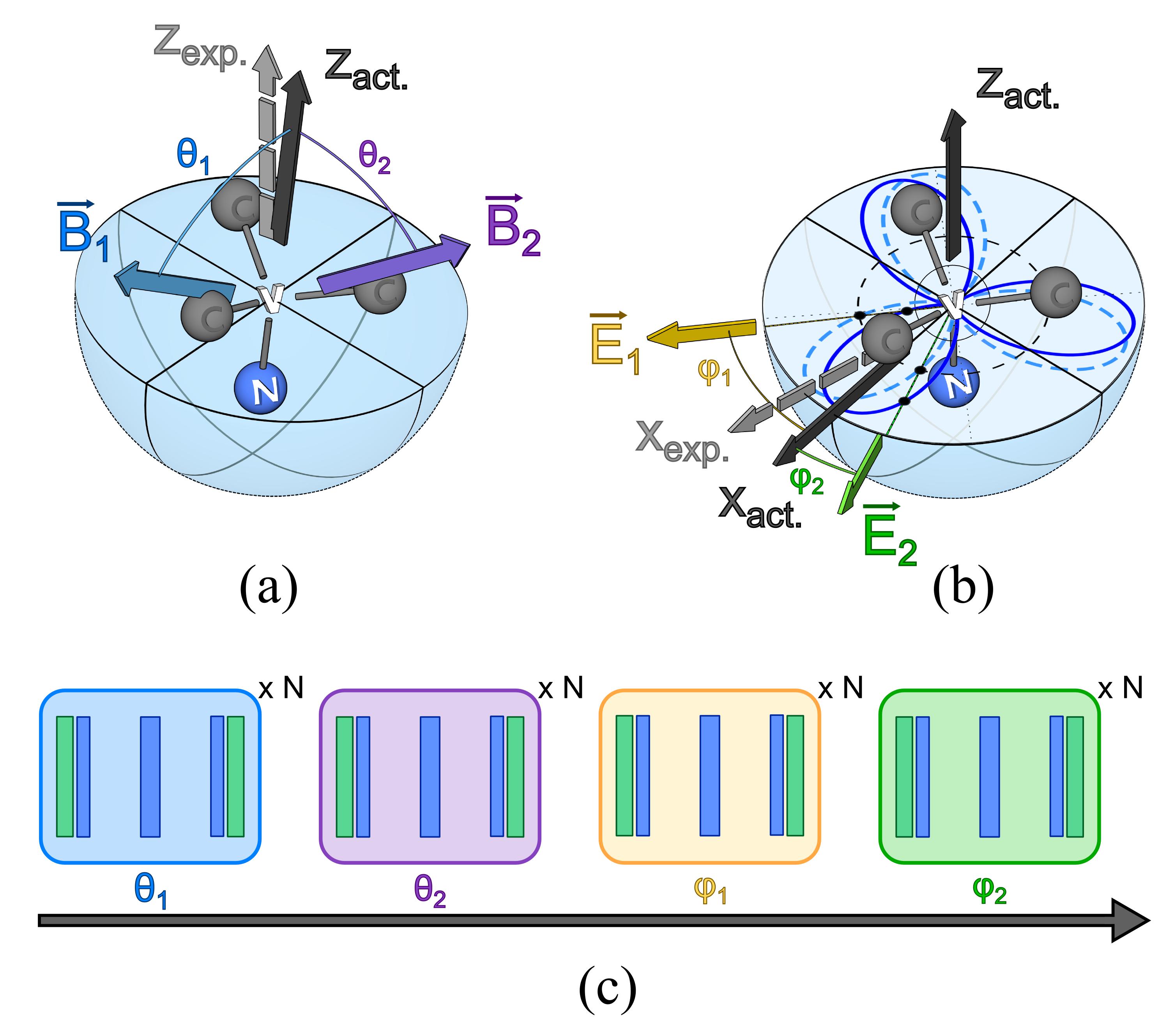}
\caption{
Proposed concept of three-axis rotation sensing.
(a) First the $z$ direction is estimated by measuring its actual $\theta_{1,2}$ angles with respect to magnetic fields $\vec{B}_{1,2}$.
The magnetic fields have an angle of $\approx 60\,\deg$ with the previous/ expected $z$ direction.
(b) Next the $x$ direction is estimated by ac-electrometry using parallel electric ($\vec{E}_{1,2}$) and magnetic (not shown) fields transverse to $z$ and with angles $\phi_{1,2}=\pm30\,\deg$ with the previous/ expected $x$ direction.
(c) Full measurement sequence for one three-axis orientation estimate.
Each angle estimate is composed of $N$ ODMR pulse sequences in order to reduce the relative photon shot noise and to improve accuracy.}
\label{fig:fig9}
\end{center}
\end{figure}

The proposed concept of three-axis rotation sensing using a single NV$^-$ centre is to employ the orientation measurement technique to measure the orientations of the centre's $x$ and $z$ coordinate axes at discrete intervals (refer to figure \ref{fig:fig9}).
From the discrete set of coordinate orientation measurements $\{\vec{x}(t_i),\vec{z}(t_i)\}$, the three-axis rotations $\{R(\beta_i,\vec{n}_i)\}$ that occurred during each time interval, such that $\vec{x}(t_{i+1})=R(\beta_i,\hat{n}_i) \cdot \vec{x}(t_i)$ and $\vec{z}(t_{i+1})=R(\beta_i,\hat{n}_i) \cdot \vec{z}(t_i)$, can be reconstructed using
\begin{eqnarray}
\vec{n}_i & = & [\vec{z}(t_{i+1})-\vec{z}(t_i)]\times[\vec{x}(t_{i+1})-\vec{x}(t_i)] \nonumber \\
\cos\beta_i & = & \vec{x}(t_{i+1})\cdot\vec{x}(t_{i})+\vec{z}(t_{i+1})\cdot\vec{z}(t_{i})+[\vec{x}(t_{i+1})\cdot\vec{x}(t_{i})][\vec{z}(t_{i+1})\cdot\vec{z}(t_{i})]\nonumber \\
&&-[\vec{x}(t_{i})\cdot\vec{z}(t_{i+1})][\vec{z}(t_{i})\cdot\vec{x}(t_{i+1})]
\end{eqnarray}
where $\vec{n}_i$ and $\beta_i$ are the axis and angle of the rotations. Note that the rotational dynamics are assumed to be slow enough such that there is no appreciable rotation during each orientation measurement. Given an apparatus where the direction of the parallel fields are precisely known with respect to a chosen reference frame, the accuracy of the orientation measurements will thus be determined by the sensitivity of the ODMR signal to misalignment of the fields.

The critical characteristic of our rotation sensing proposal is the average time $T$ required to perform an orientation measurement, which depends on the the average number of ODMR measurements for achieving a sufficient angle estimate for each axis and the relevant coherence time $T_c$ that determines the time of each ODMR measurement. NV$^-$ quantum sensing is performed via the measurement of spin frequencies using ODMR pulse sequences, such as the Ramsey, spin-echo and more advanced dynamical decoupling sequences.
In essence, during a pulse sequence a coherent superposition of two spin states (e.g. $\ket{0}$ and $\ket{-}$) acquires a phase with a certain frequency.
The latter is then sensitive to various quantities.
For a single NV center and practical experimental parameters, the frequency sensitivity is
\begin{equation}
\delta \nu \approx C \cdot T_c^{-1/2}
\label{eq:freq_sens}
\end{equation}
where $C\approx 12$ given an average fluorescence photon count rate (under cw illumination with readout laser intensity) of $200\,$kcounts, an ODMR contrast of $30\,\%$ and an exponential decay of the Ramsey or spin-echo-like signal over the phase accumulation time $\tau$.
For $\tau = T_c = 1\,$ms, this yields a frequency sensitivity of $\delta \nu=380\,\mathrm{Hz/\sqrt{Hz}}$.

For an appropriate experimental setting we can convert this frequency sensitivity into an angle sensitivity for $\theta$ or $\gamma$.
In particular, we might perform combined ac-magnetometry as depicted in fig.~\ref{fig:fig9}a (see also ref.~\cite{balasubramanian09}) and ac-electrometry as illustrated in figs.~\ref{fig:fig7} and \ref{fig:fig9}b.
In that case, we divide the frequency sensitivity $\delta\nu$ by the derivative of the frequency shift $\Delta f_-$ with respect to the angle $\theta$ or $\gamma$ (assuming a superposition of spin states $\ket{0}$ and $\ket{-}$) at the corresponding working point
\begin{eqnarray}
\label{eq:angle_sens_theta}
\delta \theta \approx \left( \left| \frac{d\left(\Delta f_-\right)}{d\theta} \right|_{\theta \approx 60\deg} \right)^{-1}  \delta \nu \\
\delta \gamma \approx \left( \left| \frac{d\left(\Delta f_-\right)}{d\gamma} \right|_{\gamma = \pm 30\deg, \delta=0} \right)^{-1}  \delta \nu
\label{eq:angle_sens_gamma}
\end{eqnarray}
Using practical field magnitudes of $B=100\,$G and $E=1\,$V$\mu$m$^{-1}$ we obtain
\begin{eqnarray}
\delta \theta \approx 8.0\cdot 10^{-5}\,\deg/\sqrt{\mathrm{Hz}} \\
\delta \gamma \approx 0.043\,\deg/\sqrt{\mathrm{Hz}} \, .
\label{eq:angle_sens_values}
\end{eqnarray}
Even higher sensitivities might be achieved when using superposition states of $\ket{-}$ and $\ket{+}$.
In that case the derivatives in equations (\ref{eq:angle_sens_theta}) and(\ref{eq:angle_sens_gamma}) do increase and we obtain for the sensitivities
\begin{eqnarray}
\delta \theta_{dq} \approx 4.5\cdot 10^{-5}\,\deg/\sqrt{\mathrm{Hz}} \\
\delta \gamma_{dq} \approx 0.021\,\deg/\sqrt{\mathrm{Hz}} \, .
\label{eq:angle_sens_values_dq}
\end{eqnarray}

In the following we describe a measurement scenario based on a rotating NV diamond and adjustable electric and magnetic fields in the lab frame (see fig.~\ref{fig:fig2}b).
In the case of a slow rotation speed compared to the measurement rate, for both the estimation of the $z$ and the $x$ direction we perform each two measurements.
Starting from the expected/ previous $z$ axis we apply magnetic fields with $\theta\approx 60\,\deg$ and two $\phi_B$ values which differ by $90\,\deg$ (see fig.~\ref{fig:fig9}a).
The result is an updated/ actual $z$ direction.
For the estimation of the $x$ axis we apply parallel electric and magnetic fields transverse to the actual $z$ direction and with angles $\phi_E=\phi_B=\pm 30\,\deg$ with respect to the expected $x$ axis (see fig.~\ref{fig:fig9}b).
Finally, $z$ and $x$ directions are updated.

If we target at a similar accuracy for $z$ and $x$ we can neglect the time for the $z$ estimation measurement compared to the $x$ estimation measurement because of the large mismatch of the respective sensitivities.
Therefore, the overall angle sensitivity for three-axis rotation sensing is equal to $\delta\gamma_{dq}\approx0.02\,\deg/\sqrt{\mathrm{Hz}}$ given the above assumptions. For a total three-axis rotation measurement time of $T=1$ s, this corresponds to an angular accuracy of 0.02 $\deg$, which compares favourably to a previous single-axis rotational sensing demonstration where the rotational dynamics of a nanodiamond within a biological cell were $\sim 1$ deg/ hour  \cite{mcguiness11}. Faster rotational dynamics can be captured at the expense of angular accuracy. For example, if a much smaller ODMR measurement time $T_c\sim 1\,\mu$s is used, the total three-axis measurement rate is increased by three orders of magnitude at the expense of a 33 fold decrease in angular accuracy. Of course, in a real three-axis rotational sensing experiment, it is important to  adjust the estimation rate to the rotational dynamics in order to exploit maximum possible sensitivities. The measurement sequence proposed here may also be improved or tailored to the specific sensing task in order to optimize sensitivity. Given these considerations, it is clear that since three-axis rotational sensing enables the rotational dynamics of a nanodiamond to be almost completely reconstructed, the technique proposed here may have significant applications in the study of the internal mechanics of biological cells.

\section{Conclusion}

In this paper, we have demonstrated an ODMR technique employing rotations of static electric and magnetic fields that precisely measures the orientation of the NV$^-$ centre's defect structure. We developed our technique through a theoretical examination of the relationship between the centre's observable spin resonances and its defect structure and showed that, by observing the variation of the spin resonances as parallel electric and magnetic fields are rotated, the fields can be aligned with specific directions within the centre's defect structure. Whilst our technique is a vital enabler of the centre's existing vector sensing applications, it also motivates new applications in multi-axis rotation sensing, NV growth characterization and diamond crystallography. Indeed, the application of our technique to multi-axis rotation sensing of nanodiamonds within biological cells has significant potential in the study of internal cellular mechanics.

\ack
The authors thank Tokuyuki Teraji and Junichi Isoya for the diamond sample used in this work. This work was supported by the ARC (DP120102232), SFB TR/21, SFB 716, Forschergruppe 1493 as well as EU projects SIQS and ERC SQUTEC and the Max Planck Society.

\appendix
\section{Spin-Hamiltonian derivation}
\setcounter{section}{1}
In this appendix, we briefly review the detailed derivation of the ground state spin-Hamiltonian (\ref{eq:spinhamiltonian}) presented in \cite{doherty12}. Through the application of perturbation theory, it was shown in \cite{doherty12} that spin-orbit and spin-spin coupling between the ground $^3A_2$ and excited $^3E$ levels results in the ground state spin becoming susceptible to electric fields, but does not result in its g-factor being significantly perturbed from the free electron value. Here, we present an abridged derivation that highlights the relationship between the spin-Hamiltonian and the defect structure of the NV$^-$ centre.

We begin with the centre's electronic Hamiltonian
\begin{eqnarray}
H_{NV} & = & H_o+V_{so}+V_{ss}+V_E+V_{B} \nonumber \\
& = & H_o+H_f
\end{eqnarray}
where $H_o$ is the purely orbital Hamiltonian that includes electronic kinetic energy and electron-nucleus and electron-electron electrostatic interactions, $V_{so}$ is spin-orbit interaction, $V_{ss}$ is spin-spin interaction, $V_E$ is the electric dipole interaction, and $V_{B}$ includes orbital and spin magnetic interactions.
Note that $V_{ss}$, $V_{E}$ and $V_{B}$ are spherically symmetric and invariant to coordinate transformations and can each be written as linear combinations of products of orbital and spin operators \cite{doherty11}.

The LS-coupling states corresponding to the triplet levels are \cite{doherty11}
\begin{eqnarray}
\Psi_{^3A_2,m_s} = \Phi_{^3A_2}\chi_{m_s}, \ \Psi_{^3E,x,m_s} = \Phi_{^3E,x}\chi_{m_s}, \ \Psi_{^3E,y,m_s} = \Phi_{^3E,y}\chi_{m_s}
\end{eqnarray}
where $\Phi_{\Gamma,j}$ are orbital wavefunctions that transform as the $j^{th}$ row of the $\Gamma$ irreducible representation of the $C_{3v}$ group and $\chi_{m_s}$ are the triplet spin states with $m_s=0,\pm1$. It is important to note that the orbital wavefunctions are defined by the centre's defect structure. The labels $x$ and $y$ recognise the fact the orbital wavefunctions $\{\Phi_{^3E,x},\Phi_{^3E,y}\}$ transform analogously to the coordinates $\{x,y\}$ under the $C_{3v}$ operations, if the coordinates are defined such that the $z$ and $x$ coordinate directions are aligned with the centre's major and (one of the three) minor symmetry axes. Hence, as in \cite{doherty12}, it is natural to adopt this coordinate system that is aligned with the centre's defect structure.

The LS-coupling states are eigenstates of $H_o$ and thus form the basis for the perturbative calculation of the fine structure interactions $H_{f}$.  Since it was shown in \cite{doherty12} that only interactions between the triplet levels perturb the ground state fine structure, the ground state spin-Hamiltonian may be immediately constructed to second-order in terms of orbital integrals as
\begin{eqnarray}
H & = & \me{\Phi_{^3A_2}}{H_f}{\Phi_{^3A_2}} \nonumber \\
&& -\frac{1}{E_o}\bra{\Phi_{^3A_2}}H_f\left(\ket{\Phi_{^3E,x}}\bra{\Phi_{^3E,x}}
+\ket{\Phi_{^3E,y}}\bra{\Phi_{^3E,y}}\right)H_f\ket{\Phi_{^3A_2}}
\end{eqnarray}
where $E_o\sim 1.945$ eV is the energy separation of the triplet levels and it is to be understood that the spin operators of $H_f$ that remain after the evaluation of the orbital integrals act upon the triplet spin states $\chi_{m_s}$. The above expression demonstrates that $H$ is $C_{3v}$ symmetric, since its behaviour under a coordinate transformation will be determined by the behaviour of the $C_{3v}$ symmetric orbital wavefunctions.

Evaluating the orbital integrals in the above yields the final expression of the spin-Hamiltonian (\ref{eq:spinhamiltonian}). Introducing the centre's defect orbitals ($a_1$, $e_x$ and $e_y$), the explicit expressions for the electric susceptibility parameters in the adopted coordinate system are \cite{doherty12}
\begin{eqnarray}
k_\perp & = &  8\sqrt{2}d_\perp D_E/E_o \nonumber \\
k_z & = & (s_{2,5}^2+s_{2,6}^2)d_z
\label{eq:electricparameters}
\end{eqnarray}
where
\begin{eqnarray}
d_\perp & = & d_x = \me{a_1(\vec{r}_1)}{e x_1}{e_x(\vec{r}_1)} \nonumber \\
d_{z} & = & \me{e_x(\vec{r}_1)}{e z_1}{e_x(\vec{r}_1)}-\me{a_1(\vec{r}_1)}{e z_1}{a_1(\vec{r}_1)}
\label{eq:dipoleintegrals}
\end{eqnarray}
are transverse and axial electric dipole moments,
\begin{eqnarray}
D_E & = & \frac{3\mu_0\mu_B^2g_e^2}{32\pi h}\bra{a_1(\vec{r}_1)e_y(\vec{r}_2)}\frac{x_{12}^2-y_{12}^2}{r_{12}^5}
\left(\ket{e_x(\vec{r}_1)e_y(\vec{r}_2)}\right. \nonumber \\
&&\left.-\ket{e_y(\vec{r}_1)e_x(\vec{r}_2)}\right)
\label{eq:spinspinintegral}
\end{eqnarray}
is an orbital integral of spin-spin interaction, $e$ is the fundamental charge, $\mu_0$ is the permeability of free space, $\vec{r}_i=x_i\hat{x}+y_i\hat{y}+z_i\hat{z}$ is the position of the $i^{th}$ electron, $\vec{r}_{12}=\vec{r}_2-\vec{r}_1= x_{12}\hat{x}+y_{12}\hat{y}+z_{12}\hat{z}$, and $s_{2,5}$ and $s_{2,6}$ are real spin-coupling coefficients formed from linear combinations of spin-orbit and spin-spin interaction integrals (refer to \cite{doherty11,doherty12} for further details).

The above demonstrates that the electric susceptibility parameters are products of the electric dipole and spin-spin interactions between the triplet levels. Consequently, the electric field interaction may be interpreted as resulting in the lowering of the symmetry of the ground state spin density. As discussed in section 2, whilst the adopted coordinate system is aligned with the centre's symmetry axes, there is still freedom to choose the precise coordinate directions so that electric susceptibility parameters are positive. Since the spin-coupling coefficients are real, and thus $s_{2,5}^2+s_{2,6}^2>0$, only the electric dipole $d_z$ determines the sign of $k_z$. Both the spin-spin and electric dipole contributions to $k_\perp$ determine its sign. However, notably the spin-spin interaction is invariant to a coordinate inversion because it is quadratic in the coordinates. It is shown in appendix B that the electric susceptibility parameters are positive for the coordinate system definition depicted in figure \ref{fig:fig1}a, where the $z$ coordinate axis is directed from the nitrogen towards the vacancy and when the $x$ coordinate axis is directed from the $z$ coordinate axis towards one of the vacancy's three nearest-neighbour carbons.

\section{Molecular orbital calculation of the electric susceptibility parameters}
\begin{figure}[hbtp]
\begin{center}
\mbox{
\subfigure[]{\includegraphics[width=0.3\columnwidth] {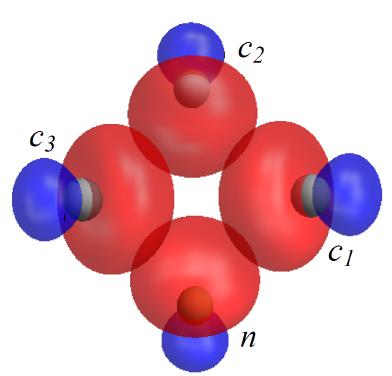}}
\subfigure[]{\includegraphics[width=0.3\columnwidth] {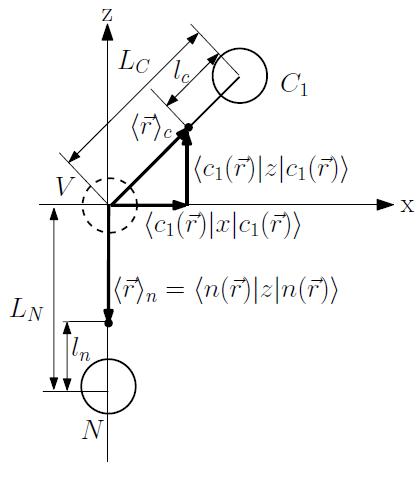}}
\subfigure[]{\includegraphics[width=0.37\columnwidth] {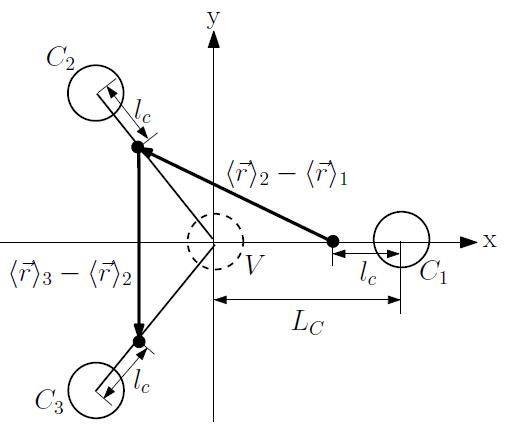}}
}
\caption{
(a) Schematic of the NV center depicting the nearest neighbor carbon atoms (gray), the substitutional nitrogen atom (brown), as well as their corresponding ($c_1$,$c_2$,$c_3$,$n$) sp$^3$ atomic orbitals (red - positive orbital contour, blue - negative orbital contour). (b) and (c) depict the geometry of the NV centre in the $xz$ and $xy$ coordinate planes, respectively. Labels are as defined in the text.}
\label{fig:appendixB}
\end{center}
\end{figure}

\setcounter{section}{2}
In this appendix, the well established molecular model of the centre \cite{doherty11,maze11} is drawn upon to perform a molecular orbital calculation of the sign of the electric susceptibility parameters. Similar calculations have been used to establish basic aspects of the temperature and pressure response of the ground state spin \cite{doherty14a,doherty14}. The molecular model exploits the highly localized nature of the centre's defect orbitals to approximate the orbitals as linear combinations of the dangling sp$^3$ atomic orbitals ($n$,$c_1$,$c_2$,$c_3$) of the vacancy's nearest neighbor nitrogen and carbon atoms (refer to figure \ref{fig:appendixB}a) \cite{doherty14a}
\begin{eqnarray}
e_x & \approx &  N_x (2c_1-c_2-c_3) \nonumber \\
e_y & \approx &  N_y (c_2-c_3)  \nonumber \\
a_1 & \approx &  N_{a_1} (c_1+c_2+c_3+\lambda n)
\end{eqnarray}
where $\lambda$ is a real linear coefficient and
\begin{eqnarray}
N_x & = & [(2\bra{c_1}-\bra{c_2}-\bra{c_3})(2\ket{c_1}-\ket{c_2}-\ket{c_3})]^{-1/2} \nonumber \\
N_y & = & [(\bra{c_2}-\bra{c_3})(\ket{c_2}-\ket{c_3})]^{-1/2} \nonumber \\
N_{a_1} & =& [(\bra{c_1}+\bra{c_2}+\bra{c_3}+\lambda\bra{n})(\ket{c_1}+\ket{c_2}+\ket{c_3}+\lambda\ket{n})]^{-1/2}
\end{eqnarray}
are normalization constants.

Expanding the electric dipole integrals (\ref{eq:dipoleintegrals}) and spin-spin integral (\ref{eq:spinspinintegral}) using the above defect orbital definitions, applying symmetry operations and ignoring orbital overlaps, the integrals become
\begin{eqnarray}
d_\perp & \approx & \frac{e}{3\sqrt{2}}\me{c_1(\vec{r})}{x}{c_1(\vec{r})} \nonumber \\
d_z & \approx & \frac{e\lambda^2}{3+\lambda^2}\left[\me{c_1(\vec{r})}{z}{c_1(\vec{r})}-\me{n(\vec{r})}{z}{n(\vec{r})}\right]\nonumber \\
D_E & \approx & \frac{3\mu_0\mu_B^2g_e^2}{32\pi h}16\frac{\sqrt{2}}{3}\left[\me{c_1(\vec{r}_1)c_2(\vec{r}_2)}{\frac{x_{12}^2-y_{12}^2}{r_{12}^5}}{c_1(\vec{r}_1)c_2(\vec{r}_2)}\right.\nonumber \\
&&\left.-\me{c_2(\vec{r}_1)c_3(\vec{r}_2)}{\frac{x_{12}^2-y_{12}^2}{r_{12}^5}}{c_2(\vec{r}_1)c_3(\vec{r}_2)}\right]
\end{eqnarray}
Using knowledge of the approximately tetrahedral nuclear geometry of the NV centre, the one-electron atomic orbital integrals within the electric dipole integrals can be evaluated. For the purpose of determining the sign of the electric dipole integrals, we can proceed with simple geometric arguments. For the carbon sp$^3$ atomic orbitals, the expected position $\langle\vec{r}\rangle_i=\me{c_i(\vec{r})}{\vec{r}}{c_i(\vec{r})}$  ($i=1,2,3$) of an electron occupying one of those orbitals occurs at a distance $l_C\sim0.31 \mathrm{\AA}$  towards the vacancy away from the carbon \cite{doherty14}. Likewise, the expected position $\langle\vec{r}\rangle_n=\me{n(\vec{r})}{\vec{r}}{n(\vec{r})}$ of an electron occupying the nitrogen orbital occurs at a distance $l_N\sim0.27 \mathrm{\AA}$ towards the vacancy away from the nitrogen. The distances between the vacancy and the carbon and nitrogen atoms are $L_C\sim1.65 \mathrm{\AA}$ and $L_N\sim1.68 \mathrm{\AA}$, respectively \cite{gali08}. Consequently, through inspection of the geometry (refer to figure \ref{fig:appendixB}b), it is clear that both electric dipole integrals are positive.

Unlike the electric dipole integrals, the spin-spin integral contains two-electron direct integrals between densities of two carbon sp$^3$ orbitals. For the purpose of determining the sign of the spin-spin integral, the difficulty of evaluating these direct integrals can be avoided and a semi-classical approximation can be performed instead, where the direct integrals between atomic orbitals are replaced by expressions containing the expected positions of the electrons occupying the orbitals \cite{doherty14}
\begin{eqnarray}
&& \me{c_1(\vec{r}_1)c_2(\vec{r}_2)}{\frac{x_{12}^2-y_{12}^2}{r_{12}^5}}{c_1(\vec{r}_1)c_2(\vec{r}_2)} \approx \nonumber \\
&&\frac{1}{|\langle\vec{r}\rangle_2-\langle\vec{r}\rangle_1|^5}
\left[(\langle x\rangle_2-\langle x\rangle_1)^2-(\langle y\rangle_2-\langle y\rangle_1)^2\right] \nonumber \\
&& \me{c_2(\vec{r}_1)c_3(\vec{r}_2)}{\frac{x_{12}^2-y_{12}^2}{r_{12}^5}}{c_2(\vec{r}_1)c_3(\vec{r}_2)} \approx \nonumber \\
&&\frac{1}{|\langle\vec{r}\rangle_3-\langle\vec{r}\rangle_2|^5}
\left[(\langle x\rangle_3-\langle x\rangle_2)^2-(\langle y\rangle_3-\langle y\rangle_2)^2\right]
\end{eqnarray}
Using these semi-classical expressions and the geometry of the centre (refer to figure \ref{fig:appendixB}c), the spin-spin integral becomes
\begin{eqnarray}
D_E & \approx & \frac{\mu_0\mu_B^2g_e^2}{4\pi h}\sqrt{\frac{2}{3}}|\me{c_1(\vec{r})}{x}{c_1(\vec{r})}|^{-3}
\end{eqnarray}
which is clearly positive. Given that each of the electric dipole and spin-spin integrals are positive, it follows from (\ref{eq:electricparameters}) that the electric susceptibility parameters are also positive for the adopted coordinate system.

\end{document}